\documentclass[sn-mathphys,Numbered,iicol]{sn-jnl}%
\usepackage{graphicx}%
\usepackage{multirow}%
\usepackage{amsmath,amssymb,amsfonts}%
\usepackage{amsthm}%
\usepackage{mathrsfs}%
\usepackage[title]{appendix}%
\usepackage{xcolor}%
\usepackage{textcomp}%
\usepackage{manyfoot}%
\usepackage{booktabs}%
\usepackage{algorithm}%
\usepackage{algorithmicx}%
\usepackage{algpseudocode}%
\usepackage{listings}%
\theoremstyle{thmstyleone}%
\theoremstyle{thmstyletwo}%

\theoremstyle{thmstylethree}%

\begin{document}

\title[Less is more: Ensemble Learning for Retinal Disease Recognition Under Limited Resources]{Less is more: Ensemble Learning for Retinal Disease Recognition Under Limited Resources}

\author[1]{\fnm{Jiahao} \sur{Wang}}\email{wangjiahaohainan@163.com}

\author[1]{\fnm{Hong} \sur{Peng}}\email{21220854000164@hainanu.edu.cn}

\author*[2]{\fnm{Shengchao} \sur{Chen}}\email{pavelchen@ieee.org}

\author*[1]{\fnm{Sufen} \sur{Ren}}\email{aug\_0815@163.com}

\affil[1]{\orgdiv{School of Information and Communication Engineering}, \orgname{Hainan University}, \orgaddress{\city{Haikou}, \postcode{570228}, \country{China}}}

\affil[2]{\orgdiv{Australian AI Institute, School of Computer Science, FEIT}, \orgname{University of Technology Sydney}, \orgaddress{\city{Sydney}, \postcode{2008}, \state{NSW}, \country{Australia}}}

%\affil[3]{\orgdiv{Department}, \orgname{Organization}, \orgaddress{\street{Street}, \city{City}, \postcode{610101}, %\state{State}, \country{Country}}}

\abstract{Retinal optical coherence tomography (OCT) images provide crucial insights into the health of the posterior ocular segment. Therefore, the advancement of automated image analysis methods is imperative to equip clinicians and researchers with quantitative data, thereby facilitating informed decision-making. The application of deep learning (DL)-based approaches has gained extensive traction for executing these analysis tasks, demonstrating remarkable performance compared to labor-intensive manual analyses. However, the acquisition of Retinal OCT images often presents challenges stemming from privacy concerns and the resource-intensive labeling procedures, which contradicts the prevailing notion that DL models necessitate substantial data volumes for achieving superior performance. Moreover, limitations in available computational resources constrain the progress of high-performance medical artificial intelligence, particularly in less developed regions and countries. This paper introduces a novel ensemble learning mechanism designed for recognizing retinal diseases under limited resources (e.g., data, computation). The mechanism leverages insights from multiple pre-trained models, facilitating the transfer and adaptation of their knowledge to Retinal OCT images. This approach establishes a robust model even when confronted with limited labeled data, eliminating the need for an extensive array of parameters, as required in learning from scratch. Comprehensive experimentation on real-world datasets demonstrates that the proposed approach can achieve superior performance in recognizing Retinal OCT images, even when dealing with exceedingly restricted labeled datasets. Furthermore, this method obviates the necessity of learning extensive-scale parameters, making it well-suited for deployment in low-resource scenarios.}

\maketitle

\section{Introduction}\label{sec1}
Visual well-being significantly impacts one's overall quality of life. Nevertheless, retinal diseases can pose substantial threats and encompass various pathological changes, such as age-related macular degeneration (AMD), diabetic macular edema (DME), and choroidal neovascularization (CNV). Specifically, AMD stands as a significant global contributor to blindness~\cite{r1}. Drusen represents a prevalent characteristic during the initial AMD phases, with its size, quantity, and morphology believed to be closely linked to the likelihood of AMD advancement~\cite{r2}. DME stands as the primary source of visual impairment among diabetic patients, characterized by retinal cysts and hardened exudates that result in retinal thickening~\cite{r3}. CNV constitutes a severe retinal ailment typically manifesting in the advanced phases of AMD~\cite{r4}. This condition can induce anomalous blood vessel growth within the macular region, thereby initiating the development of subretinal hemorrhages, leakage, and scarring. Ultimately, these factors can culminate in profound visual impairment~\cite{r5}. Although often asymptomatic during their initial phases, these diseases progressively inflict irreversible damage upon vision. As a result, ensuring early detection, expeditious and accurate diagnosis, coupled with prompt therapeutic intervention, becomes imperative in halting disease progression.

Optical coherence tomography (OCT), a non-invasive technique that can generate structural images of biological tissues by collecting optical backscattered signals, has become an important imaging modality in ophthalmology~\cite{r6,r7,r8,r9,r10}. OCT has been widely used to insight into the health of the posterior ocular segment, and to detect various retinal diseases (e.g., AMD, DME, CNV). By providing high-resolution cross-section images of ocular tissues, OCT can assist ophthalmologists in more completely evaluating the characteristics of retinal diseases. While automated image processing techniques and cross-sectional morphological analysis can quantitative thickness measurements, interpreting OCT pathology analysis results requires experienced ophthalmologists. This is an inefficient and labor-intensive manual analysis method.

With the rapid development of Artificial Intelligence (AI)~\cite{chen2022cost,chen2022fabry,chen2022reconstruction,chen2023collaborative,ren2023high,ren2023unsupervised,yuan2022efficient,xu2023dual,ren4675790distributed,chen2022high}, several Deep Learning (DL)-based computer-aided diagnostic tools~\cite{li2022improvement,wang2022semi} have been emerged to reduce the additional labor of medical personnel and improve the treatment process~\cite{r11,chen2023interpretable,chen2022cmt,peng2023diffusion}. Most of the existing work is based on the Convolutional Neural Networks (CNNs)~\cite{chen2024free,chen2023tempee,chen2022dynamic,chen2023mask} that efficiently extracting fine-grained features in OCT images and constructing their effective associations with relevant diseases~\cite{r12,r13,r14,r15,r16}. Even though they have astonishing recognition accuracy, there are several serious challenges that still limit their progress towards a more generalized road: (1) collecting a large scale of OCT images is restricted due to rising privacy concerns and potential medical device issues; (2) providing correct annotations for the OCT image dataset is an expensive and labor-intensive project that relies on the expertise of a large number of experts; (3) training a well-performing, large-scale neural network for OCT image recognition from scratch is difficult to achieve in low-resource scenarios, such as medical institutions in developing as well as less developed regions/countries. There has been a focus on achieving excellent computer-aided OCT image recognition models without resorting to expensive processes.

To address the aforementioned challenges, this paper proposes a simple yet effective ensemble learning mechanism to improve retinal disease recognition in low-resource scenarios with limited labeling. This method aims to build a reliable and high-performance recognition model by aggregating multiple pieces of prior knowledge with less annotated data for training and lower computational resource requirements. As a result, it provides more accurate analytical support for clinical diagnosis.

The main contributions of this paper are as follows:
\begin{itemize}
    \item A simple yet effective ensemble learning strategy is proposed to achieve an excellent model with less annotated training data and limited computational resources to provide accurate analytical support for clinical diagnosis from Retinal OCT images.
    \item A foundation model is proposed for the fusion of multiple prior knowledge from pre-trained models, aiming to offer an effective guide for Retinal OCT image recognition tasks. This approach avoids large-scale parameter learning and significantly reduces the requirement for annotated data.
    \item  Extensive experiments on the real-world Retinal OCT image recognition dataset demonstrate the effectiveness and superiority of our method on multiple classification evaluation metrics, which can provide promising performance under limited annotation and computational resources.
\end{itemize}

\section{Related Work}\label{sec2}

This section summarizes representative methods relevant to this paper, including Retinal OCT image recognition with CNN, Few-shot learning, and model fusion.

\subsection{Retinal OCT image recognition with CNN}\label{subsec1}
CNN is a deep learning model for image and vision tasks. Among them, ResNet~\cite{r17}, DenseNet~\cite{r18} and Inception~\cite{r19} are representative networks. They have made remarkable achievements in the field of image recognition and have made important contributions to the development of computer vision. 

The current trend in the field involves the utilization of CNN-based models to address the recognition of retinal diseases. Lee et al.\cite{r12} meticulously employed a dataset comprising 80,839 retinal OCT images to train their CNN model from inception, resulting in an impressive accuracy of 87.63\% across 20,163 validation images. Huang et al.\cite{r13} introduced a layer-guided CNN designed to identify retinal diseases. They substantiated the efficacy of their approach using two datasets featuring abundant labeled data. Mohan et al.\cite{r14} formulated a CNN architecture named MIDNet specifically for the classification of Retinal OCT images. Their model was trained using 83,484 Retinal OCT images, culminating in an impressive accuracy of 98.86\% across 968 test images. Similarly, Sunija et al.\cite{r15} proposed a lightweight CNN tailored for the classification of retinal diseases. The model underwent training and validation processes employing a dataset comprising 83,484 retinal OCT images, resulting in a remarkable accuracy of 99.69\%. Thomas et al.~\cite{r16} devised a multipath CNN aimed at achieving precise diagnosis of AMD. Their model, after being trained on a set of 87,264 retinal OCT images, exhibited elevated classification performance. This notable performance of the aforementioned CNN models is attributable to the utilization of a substantial volume of labeled data. However, access to medical image data is frequently impeded by privacy concerns and other factors. Furthermore, training a model from scratch that incorporates an extensive array of parameters inherently demands a substantial allocation of computational resources. This circumstance exacerbates the challenges of model training when computational resources are limited.

\subsection{Few-shot Learning}\label{subsec2}
Few-shot learning leverages a limited training dataset to enable the model to achieve commendable performance despite data scarcity~\cite{chen2023foundation,chen2023spatial,chen2023prompt}. Lee et al.~\cite{r20} introduced a pre-training technique that bolsters the feature extractor via contrastive learning, a method highly applicable to few-shot learning scenarios. Chen et al.~\cite{r21} applied knowledge transfer from a many-shot dataset to a novel few-shot task, yielding dependable classification performance. Zheng et al.~\cite{r22} devised a cooperative density loss module to optimize feature formation for enhanced discrimination in few-shot learning. Kim et al.~\cite{r23} introduced a multi-scale feature fusion network that targets the extraction of valuable texture and linguistic features from constrained image data. Zhu et al.~\cite{r24} formulated a dual-layer learning mechanism guided by attributes, aiming to attain more discriminative representations for enhanced classification in few-shot learning. Lim et al.~\cite{r25} employed a pre-trained model as a feature extractor to acquire valuable features in the context of few-shot learning.

In this study, we address challenges such as privacy protocol constraints related to patient data in real-world hospital settings, the resource-intensive nature of annotating extensive data, and limited computational resources. Drawing inspiration from the aforementioned studies, we integrate the concept of few-shot learning into our methodology to construct dependable models tailored for scenarios characterized by scarce labeled data and constrained computational resources.

\subsection{Ensemble Strategy}\label{subsec3}
Ensemble represents a unified learning technique designed to enhance performance by amalgamating multiple distinct underlying models. Wei et al.~\cite{r26} further advanced classification performance through the fusion of DenseNet and ConvNeXt models, incorporating deep and shallow feature fusion, and introducing an attention module. Li et al.~\cite{r27} introduced a three-backbone network model and enhanced classification performance by fusing models such as VGG16, DenseNet121, and ResNet50. Liu et al.~\cite{r28} elevated classification accuracy by fusing models including EfficientNetv2, Vision Transformer, and DenseNet. Ai et al.~\cite{r29} employed three distinct fusion strategies to amalgamate models like Inception V3, Inception-ResNet, and Xception, thereby enhancing prediction accuracy. Latha et al.~\cite{r30} enhanced the classification performance of their model by sequentially integrating features extracted from the Inception V3 and VGG16 models. Nevertheless, the fusion of models in the aforementioned studies led to an augmentation in the count of trainable parameters, subsequently intensifying the computational resource requirements during model training. This circumstance curtails the viability of model fusion in scenarios constrained by resources, particularly when multiple models are amalgamated.

\section{Methodology}\label{sec3}
In this section, we elaborate on our proposed ensemble learning mechanism for retinal OCT image classification under limited annotated data and computational resources. 
\subsection{Multi-Source Knowledge Fusion}\label{subsec4}
\begin{figure}[ht]
\centering
\includegraphics[width=0.5\textwidth]{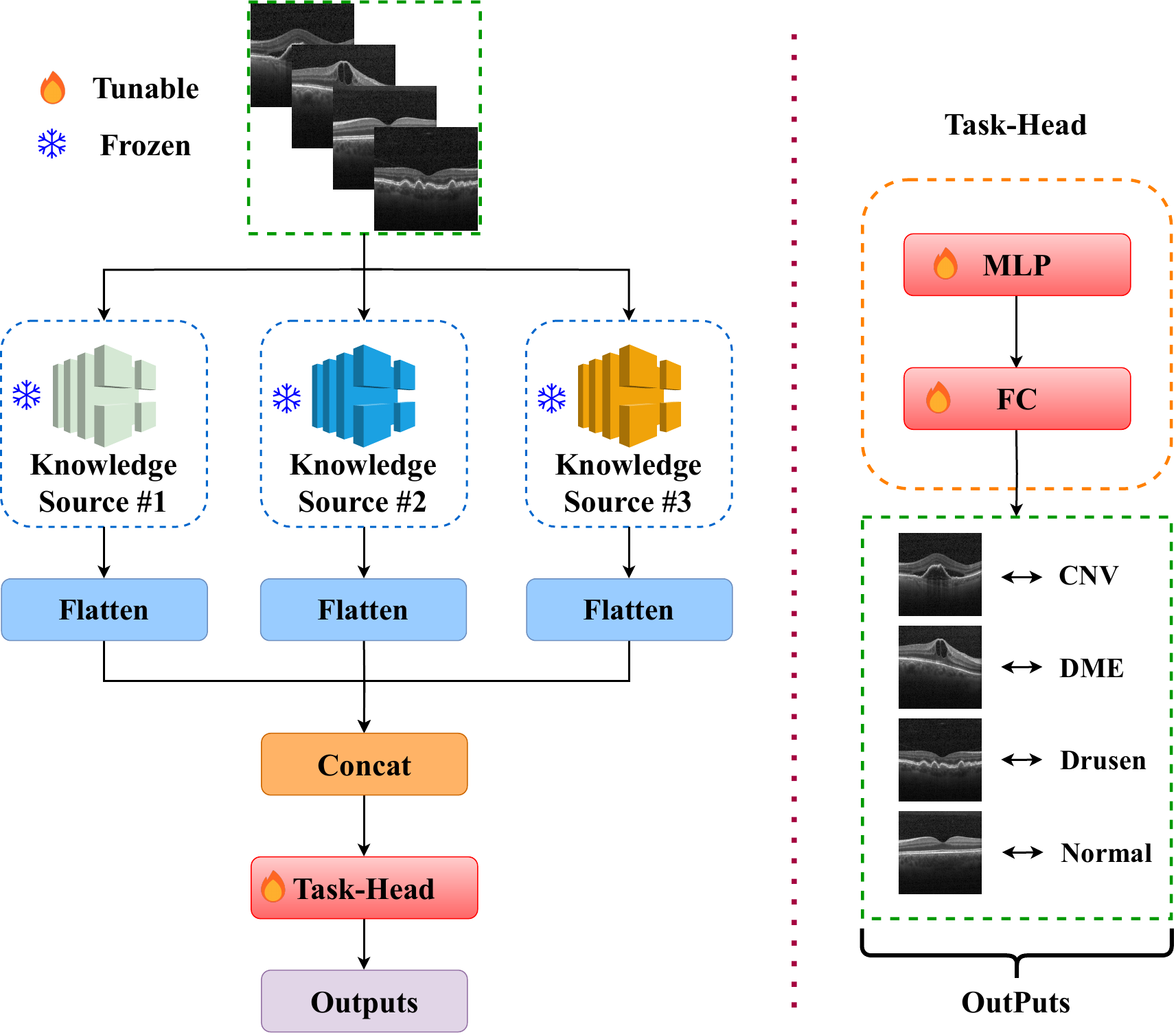}
\caption{The model architecture constructed based on the multi-source knowledge fusion strategy.}\label{fig1}
\end{figure}

The central element of our proposed approach is multi-source knowledge fusion, which enhances the flexibility and performance of neural networks in downstream tasks by leveraging prior knowledge from diverse domains. The model architecture, illustrated in Figure~\ref{fig1}, is constructed based on the multi-source knowledge fusion strategy. This architecture encompasses various pre-trained models sourced from distinct image domains, accompanied by a task-specific head. These pre-trained models remain fixed during training, effectively lowering computational expenses. Simultaneously, they contribute substantial prior knowledge, bridging not only across multiple image domains but also among disparate models. The employed pre-trained models include ResNet18, DenseNet121, and Inception V3, all of which have been pre-trained on the ImageNet dataset~\cite{r31}.

To mitigate the impact of cross-domain feature discrepancies between Retinal images and ImageNet\footnote{A large-scale image dataset comprising real-life images}, and to amalgamate the extracted representations from pre-trained models, a task-specific head is employed. This head fine-tunes the models and facilitates accurate Retinal image recognition predictions. Detailed information about the feature extractor, encompassing the aforementioned pre-trained models and the task-specific head, is provided below.

\subsection{Pre-trained Model}\label{subsec5}

ResNet18, DenseNet121, and Inception V3 are representative networks in the field of image recognition, where ResNet18 solves the problem of gradient vanishing in deep networks by introducing residual blocks; DenseNet121 constructs the network in such a way that each layer is densely connected to every other layer to promote feature reuse; Inception V3 uses multi-scale feature extraction and connectivity approaches for multi-layer feature capture. These pre-trained models have been trained on the large-scale image dataset and learned rich representations of image features, which can provide the model based on ensemble method substantial prior knowledge.

\subsection{Feature Extractor}\label{subsec6}

To extract features from retinal OCT images, we use the feature extraction layers of the ResNet18, DenseNet121, and Inception V3 pre-trained models as the base feature extractors. Specifically, we removed the heads of these pre-trained models and froze the rest of the trainable parameters. Therefore, these parameters do not need to be involved in training, effectively reducing the computational cost.

\subsection{Task Head and Knowledge Fusion}\label{subsec7}

\begin{figure}[ht]
\centering
\includegraphics[width=.5\textwidth]{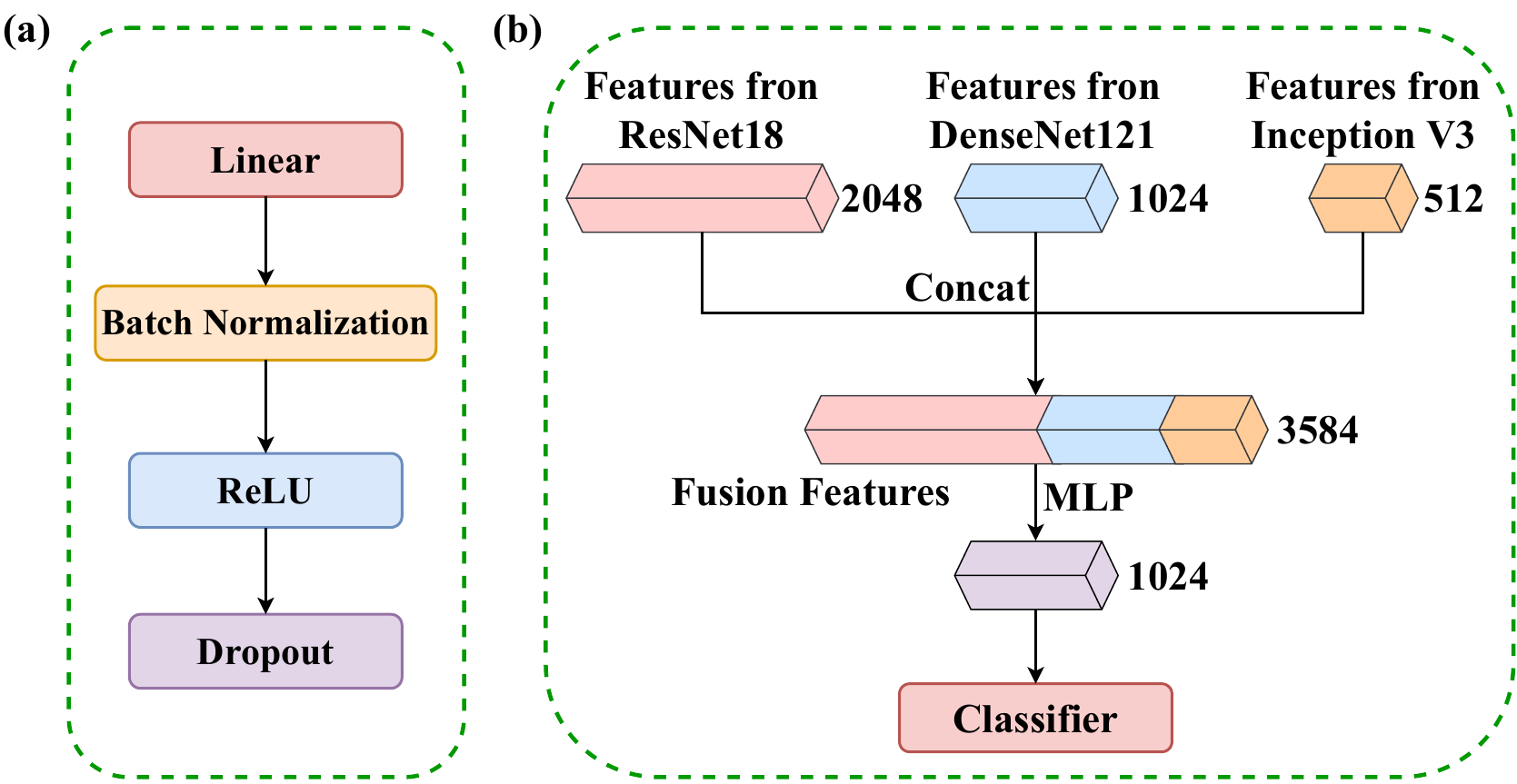}
\caption{The structure of the MLP and the process of knowledge transfer and adaptation. (a) The structure of the MLP; (b) The process of knowledge transfer and adaptation.}\label{fig2}
\end{figure}

To keep the task head simple and efficient, the task header used consists of a Multi-Layer Perceptron (MLP) and a Fully Connected Layer (FC) cascade. Among them, the MLP consists of a Linear layer, a Batch Normalization layer, a ReLU layer, and a Dropout layer cascade together. The Batch Normalization layer serves to normalize the inputs of each batch during the training process to accelerate convergence and prevent the gradient vanishing or gradient explosion problem, the Dropout layer serves to randomly discard some neurons during the training process to prevent the overfitting problem, and the ReLU layer is a nonlinear activation function that is used to introduce nonlinear properties. The ReLu activation function is formulated as:
\begin{equation}
\operatorname{ReLU}(x)=\max (x, 0)=\left\{
\begin{array}{cc}
0, & x<0 \\
 x, & x>0
\end{array}\right\}.
\label{eq1}
\end{equation}
The role of the MLP is to further process the features extracted by base feature extractors in a non-linear manner to learn the complex relationships from the features, which in turn enhances the representation of the features. Eventually, the FC layer is used as a classifier to establish relationships between the processed features and retinal disease information. Thus, the whole task head can effectively prompt the transfer and adaptation of prior knowledge from multiple pre-trained models to the task of recognizing retinal OCT images. The structure of the MLP is shown in Fig.~\ref{fig2}(a), and the process of knowledge transfer and adaptation is shown in Fig.~\ref{fig2}(b).

\subsection{Optimiser and Loss Function}\label{subsec8}
In classification tasks, achieving high performance hinges on the model's training process. One pivotal factor influencing training quality is the selection of an appropriate optimizer. To enhance the learning of image features and improve performance throughout training, we opted for the Adam as the optimizer. Adam, known for its adaptive learning rate, amalgamates the strengths of momentum gradient descent and adaptive learning rates. This enables it to dynamically modulate the learning rate across distinct parameters, thus effectively updating model parameters~\cite{r32}. The algorithm can be formulated as follows:

\begin{equation}
\begin{aligned}
& m_t=\beta_1 * m_{t-1}+(1-\beta_1) * g_t, \\
& n_t=\beta_2 * n_{t-1}+(1-\beta_2) * g_t^2, \\
& \widehat{m_t}=\frac{m_t}{1-\beta_1^t}, \\
& \widehat{n}_t=\frac{n_t}{1-\beta_2^t}, \\
& \theta_t=\theta_{t-1}-\alpha * \frac{\widehat{m_t}}{\sqrt{\widehat{n}_t}+\varepsilon},
\end{aligned}
\label{eq2}
\end{equation}
where $t$ denotes the current batch, $m_t$ and $n_t$ denote the first and second order moment momentum of the gradient, $\beta_1$ and $\beta_2$ denote the decay rates of first and second order moment momentum, $\widehat{m_t}$ and $\widehat{n_t}$ are deviation corrections to $m_t$ and $n_t$, $\theta_t$ denotes the parameters of the current batch model, $\alpha$ denotes learning rate, and $\varepsilon$ is a very small constant used to prevent the denominator from going to zero.

Retinal disease recognition is a multi-class classification task. We utilize the cross-entropy as the loss function to evaluate the model performance during the training process, which is formulated as follows:
\begin{equation}
Loss=-\sum_{i=1}^N q_i \log \left(p_i\right),
\label{eq3}
\end{equation}
where $N$ is the number of classification categories, $q$ denotes the true distribution of the sample, and $p$ presents the distribution predicted by the model.

\section{Experiments}\label{sec4}

In this study, the experimental setup encompasses the utilization of the Windows 11 operating system, and an NVIDIA GeForce RTX 3060 GPU with 12 GB of VRAM. All experimental procedures were executed utilizing Python 3.8.5 and PyTorch 1.8.0. The neural network optimization employed the Adam optimizer with an initial momentum of 0.9, a secondary momentum of 0.999, and an initial learning rate of 0.0001. The training process involved subjecting all models to 50 epochs each.

\subsection{Dataset}\label{subsec9}
The dataset utilized in this study comprises OCT images representing distinct retinal symptoms: CNV, DME, Drusen, and Normal. The selection was performed by Heidelberg Engineering using retrospective cohorts of adult patients sourced from various institutions, including the Shiley Eye Institute at the University of California, San Diego, and the Beijing Tongren Eye Center. This dataset consists of a total of 84,495 images, each meticulously validated and corrected through multiple layers of trained graders.

\begin{figure}[ht]
\centering
\includegraphics[width=0.485\textwidth]{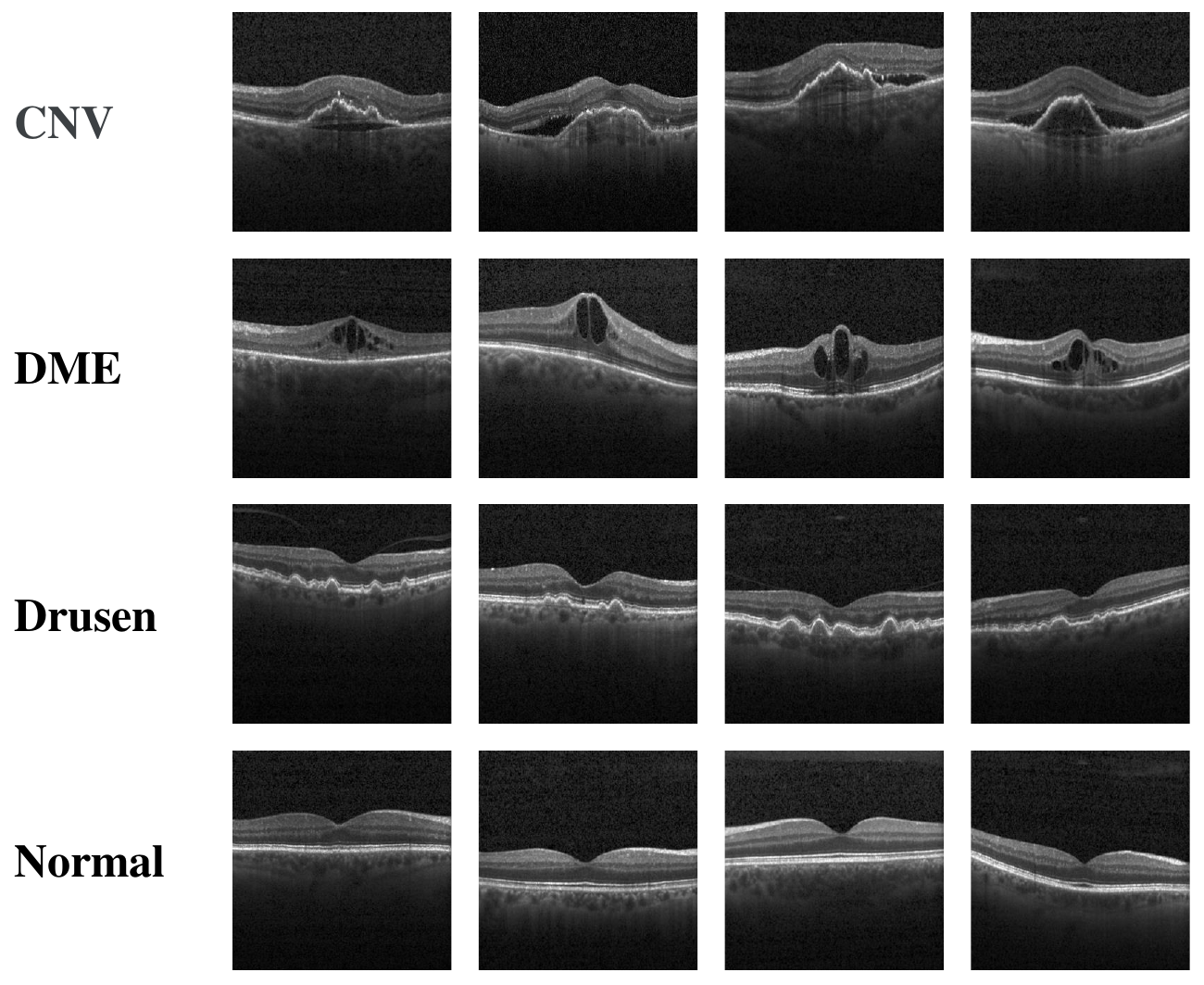}
\caption{Examples of the OCT image dataset used in this work, from top to bottom, are CNV, DME, Drusen, and Normal.}\label{fig3}
\end{figure}

To assess the classification performance of the models under conditions of limited labeled data, we gathered a random sample of 10,000 images from this dataset. Subsequently, we designated 500 images for the training set, 1,500 for validation, and 8,000 for testing. Additionally, to align with the three-channel input requirement of the pre-trained model, we conducted channel expansion on the input images. This involved duplicating gray values from a single channel into three channels and resizing all images to dimensions of 224 × 224. Detailed dataset distribution is presented in Table 1, while Fig.~\ref{fig3} visually depicts a selection of OCT images from the dataset.

\begin{table*}[ht]
\caption{Specific distribution of the dataset.}\label{tab1}
\begin{tabular}{@{}cccccc@{}}
\toprule
\textbf{Number} & \textbf{Training}& \textbf{Validation} & \textbf{Testing} & \textbf{Resolution} &\textbf{Total}\\
\midrule
CNV     &119  &357  &1904  &224*224  &2380  \\
DME	    &122  &364  &1944  &224*224  &2430  \\
Drusen  &121  &365  &1944  &224*224  &2430  \\
Normal  &138  &414  &2208  &224*224  &2760  \\
Total   &500  &1500 &8000  &  --     &10000 \\
\bottomrule
\end{tabular}
\end{table*}

\subsection{Evaluation Metrics}\label{subsec10}
To assess the classification performance of the models, we adopt widely recognized evaluation metrics for multi-class classification tasks: Precision, Recall, F1-score, And accuracy. Precision represents the ratio of samples predicted to belong to a certain category that indeed to that category. It quantifies the model's capacity for accurate category classification. In contrast, Recall gauges the proportion of samples belonging to a particular category that are accurately predicted within that category. It reflects the model's competency in category classification. The F1-score harmonizes Precision and Recall into a single metric, providing a comprehensive assessment of the model's classification performance. Accuracy, a prevalent metric, signifies the ratio of correctly classified samples to the total samples, offering insight into the model's effectiveness. A higher accuracy denotes superior classification performance. The formulas for Precision, Recall, F1-score, and Accuracy are formulated as follows:
\begin{equation}
    \text { Precision }=\frac{T P}{T P+F P},
\label{eq4}
\end{equation}

\begin{equation}
    \text { Recall }=\frac{T P}{T P+F N},
\label{eq5}
\end{equation}

\begin{equation}
    \text { F1-score }=\frac{2T P}{2T P+F P+F N},
\label{eq6}
\end{equation}

\begin{equation}
    \text { Accuracy }=\frac{T P+T N}{T P+T N+F P+F N},
\label{eq7}
\end{equation}
Where, the terms True Positive (TP), True Negative (TN), False Positive (FP), and False Negative (FN) hold specific and vital meanings. TP indicates instances where the model accurately predicts a sample as belonging to a particular category. TN, on the other hand, signifies the model's precise prediction of samples as belonging to categories other than the one under consideration. Conversely, FP denotes the model's inaccurate prediction of samples as belonging to a specific category when, in fact, they do not. Lastly, FN indicates situations in which the model incorrectly predicts samples from one category to be members of other categories.

\subsection{Performance Comparsion}\label{subsec11}

In our setting, ResNet18, DenseNet121, and Inception V3, pre-trained by ImageNet, were selected as the baseline models for comparison. Moreover, pairwise ensemble models, constructed by the proposed ensemble method, were also used for comparison to further validate the effectiveness of the proposed method. To evaluate the performance of the models, Precision, Recall, F1 score, and Accuracy were calculated for each model on the test set. The experimental results are shown in Table 2.

As can be seen from Table 2, the model based on the proposed ensemble method and pairwise ensemble models outperforms the baseline models, although the baseline models have achieved reliable performance. Among them, the model based on the proposed ensemble method achieves the highest accuracy of 92.06\% (higher 7.04\%, 6.81\%, and 9.23\% than ResNet18, DenseNet121, and Inception V3 base models, respectively). This means that the knowledge of pre-trained models can be transferred and adapted to the downstream task of Retinal OCT image recognition, and the proposed method can promote the fusion of the knowledge from multiple pre-trained models to achieve better performance. Combining the results of precision, Recall, and F1 score, the model based on the proposed ensemble method achieves state-of-the-art results on this dataset. This indicates that models constructed according to the proposed ensemble approach can obtain promising retinal disease recognition performance under limited annotated data.

\begin{table*}[ht]
\centering
\caption{The experimental results of 500 retinal OCT images as the training set, with ×100 for all results relative to the initial results, bold represents the best value.}\label{tab2}
\begin{tabular}{@{}cccccc@{}}
\toprule
\textbf{Model} & \textbf{Classes}& \textbf{Precision} & \textbf{Recall} & \textbf{F1-score} &\textbf{Accuracy}\\
\midrule
\multirow{4}{*}{\text {ResNet18}}  & CNV    & 82.48 & 80.36 & 81.40 &\multirow{4}{*} {85.02} \\
                                   & DME    & 86.81 & 82.30 & 84.50 \\
                                   & Drusen & 80.58 & 81.74 & 81.15 \\
                                   & Normal & 89.40 & 94.34 & 91.80 \\ 
\hline
\multirow{4}{*}{\text {DenseNet121}}  & CNV    & 81.38 & 83.09 & 82.22 &\multirow{4}{*} {85.25} \\
                                      & DME    & 86.20 & 86.11 & 86.16 \\
                                      & Drusen & 85.28 & 76.90 & 80.88 \\
                                      & Normal & 87.63 & 93.70 & 90.57 \\ 
\hline
\multirow{4}{*}{\text {Inception V3}}  & CNV    & 78.14 & 72.27 & 75.09 &\multirow{4}{*} {82.83} \\
                                       & DME    & 82.02 & 81.89 & 81.96 \\
                                       & Drusen & 78.25 & 82.36 & 80.25 \\
                                       & Normal & 91.34 & 93.16 & 92.24 \\ 
\hline
\multirow{4}{*}{\text {ResNet18+DenseNet121}}  & CNV    & \textbf{90.67} & 82.67 & 86.48 &\multirow{4}{*} {90.21} \\
                                               & DME    & 91.55 & 90.84 & 91.20 \\
                                               & Drusen & 85.59 & 89.56 & 87.53 \\
                                               & Normal & 92.83 & 96.74 & 94.74 \\ 
\hline
\multirow{4}{*}{\text {ResNet18+Inception V3}}  & CNV    & 84.82 & 85.14 & 84.98 &\multirow{4}{*} {89.15} \\
                                                & DME    & 93.93 & 85.96 & 89.77 \\
                                                & Drusen & 84.96 & 87.76 & 86.34 \\
                                                & Normal & 92.70 & 96.65 & 94.63 \\ 
\hline
\multirow{4}{*}{\text {DenseNet121+Inception V3}}  & CNV    & 87.09 & 81.14 & 84.01 &\multirow{4}{*} {89.31} \\
                                                   & DME    & 90.43 & 89.45 & 89.94 \\
                                                   & Drusen & 84.45 & 89.66 & 86.98 \\
                                                   & Normal & 94.60 & 95.92 & 95.26 \\ 
\hline
\multirow{4}{*}{\text {Ours}}  & CNV    & 90.45  & \textbf{86.55} & \textbf{88.46} &\multirow{4}{*} {\textbf{92.06}} \\
                                                & DME    & \textbf{94.54} & \textbf{91.72} & \textbf{93.11} \\
                                                & Drusen & \textbf{87.84} & \textbf{91.77} & \textbf{89.76} \\
                                                & Normal & \textbf{95.09} & \textbf{97.37} & \textbf{96.22} \\
\bottomrule
\end{tabular}
\end{table*}

\subsection{Resource Consumption \& Parameters Utilization}
The trainable parameters comparison results of the model based on the proposed ensemble method and the baseline models are presented in Table 3. The proposed method freezes the trainable parameters of the pre-trained models so that only a small number of parameters in the task head need to be fine-tuned during the model training process, which significantly reduces the trainable parameters of the model during the training process from the original $43.593M$ to $3.677M$. In addition, the trainable parameters of the model based on the proposed ensemble method are reduced by 68.54\%, 53.91\%, and 84.57\% compared to the ResNet18, DenseNet121, and Inception V3 baseline models, respectively. This indicates that the proposed ensemble method can effectively reduce the demand for computational resources.

\begin{table*}[ht]
\centering
\caption{Comparison results of trainable parameters between the model based on the proposed ensemble method and the baseline models, where Learn from scratch indicates that no pre-trained parameters are used, and Pre-trained indicates that pre-trained parameters are used and frozen, bold represents the best value.}\label{tab3}
\begin{tabular}{ccc}
\toprule
\textbf{Model} & \textbf{Training mode}& \textbf{Trainable parameter}\\
\midrule
ResNet18      &\text{Learn from scratch}  &\text{11.690~M}\\
DenseNet121	  &\text{Learn from scratch}  &\text{7.979~M}\\
Inception V3  &\text{Learn from scratch}  &\text{23.835~M}\\
\hline
\multirow{2}{*}{\text {Ours}}   &\text{Learn from scratch}  &\text{43.593~M}\\
                                &\text{Pre-trained}         &\textbf{3.677~M}\\
\bottomrule
\end{tabular}
\end{table*}

To further explore the potential of the model based on the proposed ensemble method under extremely limited data, we reduced the number of training samples. From the collected dataset, 20, 30, and 50 images of each class are randomly selected as the training set, and the remaining images are divided into validation and test sets in a ratio of 1:9. The experimental results are shown in Table 4.

\begin{table*}[ht]
\centering
\setlength{\tabcolsep}{1pt}
\caption{The experimental results of 20, 30, and 50 retinal OCT images per class as the training set, with ×100 for all results relative to the initial results, bold represents the best value.}\label{tab4}
\scalebox{0.7}{
\begin{tabular}{@{}cccccccccccccc@{}}
\toprule
& \multicolumn{4}{@{}c@{}}{20-shot} & \multicolumn{4}{@{}c@{}}{30-shot} & \multicolumn{4}{@{}c@{}}{50-shot}\\\cmidrule{3-6}\cmidrule{7-10}\cmidrule{11-14}%
Model &Classes & Precision & Recall & F1-score & Accuracy & Precision & Recall & F1-score & Accuracy & Precision & Recall & F1-score & Accuracy\\
\midrule
\multirow{4}{*}{\text {ResNet18}}  & CNV    & 74.93 & 65.87 & 70.11 &\multirow{4}{*} {73.70} & 69.81 & 68.32 & 69.06 &\multirow{4}{*} {75.62} & 81.41 & 77.06 & 79.18 &\multirow{4}{*} {82.18} \\
                                   & DME    & 70.85 & 77.64 & 74.09 && 75.74 & 72.69 & 74.18 && 80.07 & 78.94 & 79.50\\
                                   & Drusen & 65.11 & 73.21 & 68.92 && 69.95 & 67.45 & 68.68 && 76.62 & 81.23 & 78.86\\
                                   & Normal & 85.03 & 77.41 & 81.04 && 84.47 & 91.66 & 87.92 && 89.76 & 90.24 & 90.00\\ 
\hline
\multirow{4}{*}{\text {DenseNet121}}  & CNV    & 66.64 & 71.94 & 69.19 &\multirow{4}{*} {73.34}   & 76.18 & 72.72 & 74.41 &\multirow{4}{*} {79.23} & 80.72 & 78.45 & 79.56 &\multirow{4}{*} {81.93} \\
                                      & DME    & 74.16 & 73.31 & 73.73 && 78.20 & 77.22 & 77.71 && 83.23 & 80.86 & 82.03\\
                                      & Drusen & 70.32 & 64.45 & 67.26 && 72.59 & 80.93 & 76.53 && 79.67 & 73.02 & 76.20\\
                                      & Normal & 81.18 & 82.40 & 81.79 && 89.67 & 85.10 & 87.33 && 83.46 & 93.69 & 88.28\\
\hline
\multirow{4}{*}{\text {Inception V3}}  & CNV    & 57.62 & 60.31 & 58.94 &\multirow{4}{*} {68.59}   & 65.71 & 59.34 & 62.36 &\multirow{4}{*} {73.08} & 72.44 & 63.81 & 67.85 &\multirow{4}{*} {76.03} \\
                                       & DME    & 64.98 & 63.99 & 64.48 && 71.46 & 70.69 & 71.07 && 73.63 & 73.25 & 73.44\\
                                       & Drusen & 69.33 & 58.46 & 63.43 && 68.82 & 68.06 & 68.44 && 68.99 & 74.28 & 71.54\\
                                       & Normal & 79.82 & 88.69 & 84.02 && 82.91 & 91.41 & 86.95 && 87.07 & 90.53 & 88.76\\
\hline
\multirow{4}{*}{\text {ResNet18+DenseNet121}}  & CNV    & 79.30 & \textbf{76.27} & \textbf{77.75} &\multirow{4}{*} {79.16}   & 80.52 & 75.65 & 78.01 &\multirow{4}{*} {84.02} & 85.43 & 80.83 & 83.07 &\multirow{4}{*} {85.63} \\
                                               & DME    & \textbf{80.93} & 81.60 & 81.27 && 84.02 & 83.24 & 83.63 && 87.28 & 82.68 & 84.92\\
                                               & Drusen & 72.47 & 74.50 & 73.47 && 78.44 & 82.18 & 80.26 && 83.29 & 80.77 & 82.01\\
                                               & Normal & 83.51 & 83.58 & 83.54 && 91.85 & 93.53 & 92.68 && 86.30 & 96.60 & 91.16\\
\hline
\multirow{4}{*}{\text {ResNet18+Inception V3}} & CNV    & 72.18 & 75.61 & 73.86 &\multirow{4}{*} {77.99}   & 74.10 & 72.91 & 73.50 &\multirow{4}{*} {80.89} & 83.03 & 80.97 & 81.99 &\multirow{4}{*} {86.01} \\
                                               & DME    & 80.06 & 78.65 & 79.35 && 82.02 & 78.75 & 80.35 && 86.07 & 84.50 & 85.28\\
                                               & Drusen & 72.24 & 70.77 & 71.50 && 76.21 & 75.05 & 75.62 && \textbf{84.46} & 80.21 & 82.28\\
                                               & Normal & 86.47 & 85.81 & 86.14 && 89.23 & 94.79 & 91.93 &&89.46  & \textbf{96.76} & 92.97\\
\hline
\multirow{4}{*}{\text {DenseNet121+Inception V3}} & CNV    & 75.28 & 67.23 & 71.03 &\multirow{4}{*} {77.69}   & 78.55 & 73.76 & 76.08 &\multirow{4}{*} {82.42} & 82.31 & 80.97 & 81.63 &\multirow{4}{*} {85.50} \\
                                                  & DME    & 77.87 & 77.69 & 77.78 && 80.78 & 82.13 & 81.45 && 88.39 & 81.37 & 84.74\\
                                                  & Drusen & 72.23 & 76.16 & 74.15 && 78.00 & 79.63 & 78.81 && 82.59 & 81.51 & 82.05\\
                                                  & Normal & 84.15 & 88.04 & 86.05 && 90.82 & 92.59 & 91.70 && 88.13 & 96.51 & 92.13\\
\hline
\multirow{4}{*}{\text {Ours}} & CNV    & \textbf{81.89} & 70.24 & 75.62 &\multirow{4}{*} {\textbf{81.31}} & \textbf{81.22} & \textbf{77.49} & \textbf{79.31} &\multirow{4}{*} {\textbf{85.84}} & \textbf{87.98} & \textbf{81.69} & \textbf{84.72} &\multirow{4}{*} {\textbf{88.07}} \\
                                               & DME    & 77.69 & \textbf{86.86} & \textbf{82.02} && \textbf{85.03} & \textbf{86.53} & \textbf{85.77} && \textbf{89.45} & \textbf{86.65} & \textbf{88.02}\\
                                               & Drusen & \textbf{76.28} & \textbf{79.90} & \textbf{78.05} && \textbf{82.78} & \textbf{82.59} & \textbf{82.69} && 84.18 & \textbf{85.95} & \textbf{85.05}\\
                                               & Normal & \textbf{89.25} & \textbf{87.19} & \textbf{88.21} && \textbf{92.86} & \textbf{95.28} & \textbf{94.05} && \textbf{90.31} & 96.68 & \textbf{93.39}\\
        \bottomrule
\end{tabular}}
\end{table*}

As can be seen in Table 4, the accuracy of both the model based on the proposed ensemble method and the pairwise ensemble models improved compared to the baseline models under the three extremely limited labeled data. Among them, the model based on the proposed ensemble method all obtained the highest accuracy (81.31\%, 85.84\%, and 88.07\% in the experiments of 20, 30, and 50 retinal OCT images per class as the training set). This implies that even when dealing with extremely limited retinal OCT image data, the proposed ensemble method can well fuse the knowledge of the pre-trained models and make the knowledge transferable and adaptable to the recognition task of retinal OCT images. In terms of precision, Recall, and F1-score, the model based on the proposed ensemble method is also further improved, especially in the experiment with 30 images per class as the training set. The model achieves the best values for all metrics. This indicates that the proposed ensemble approach can build robust and reliable models even when confronted with extremely limited retinal OCT image data.

\section{Conclusion and Future Work}\label{sec5}
In conclusion, this paper proposes a simple yet effective ensemble learning mechanism to address the challenge of retinal disease recognition under sparse labeled data. By adequately integrating the prior knowledge of multiple pre-trained models, which effectively improves the knowledge transfer and adaptation of the models on retinal OCT images. The method provides effective guidance and avoids large-scale parameter learning, despite the constraints of limited labeled datasets. Extensive experimental results show that the method performs excellently on the task of processing retinal OCT images, and provides precise and reliable recognition models for disease diagnosis with limited labeled datasets. Moreover, this research also provides a practical way to solve the medical image analysis task under the problem of resource scarcity, which has the potential to promote medical applications.

In future work, we will try to ensemble more pre-trained models under sparse data and explore other types of medical image data. The aim is to develop rapid and accurate DL models for the medical field in the reality of insufficient medical data and limited computational resources.

\bibliography{ref}

\end{document}